\documentclass[conference]{IEEEtran}
\IEEEoverridecommandlockouts
\usepackage{amsmath,amssymb,amsfonts}
\usepackage{algorithmic}
\usepackage{graphicx}
\usepackage{textcomp}
\usepackage{xcolor}
\usepackage{float}
\usepackage{paralist}
\usepackage{hyperref} 
\usepackage[compact]{titlesec}
\titlespacing{\section}{0pt}{1ex}{0ex}
\titlespacing{\subsection}{0pt}{0.1ex}{0ex}\titlespacing{\subsubsection}{0pt}{0.1ex}{0ex}

\usepackage[backend=bibtex,style=ieee]{biblatex}
\begin{document}

\title{Home Energy Management Systems: Challenges, Heterogeneity \& Integration Architecture Towards A Smart City Ecosystem\\}

\author{%
\IEEEauthorblockN{Georgios Kormpakis\IEEEauthorrefmark{1}, Alexios Lekidis\IEEEauthorrefmark{1}\IEEEauthorrefmark{3}, Elissaios Sarmas\IEEEauthorrefmark{1}, Giannis Papias\IEEEauthorrefmark{1},\\ Filippos Serepas \IEEEauthorrefmark{2}, George Stravodimos \IEEEauthorrefmark{2}, Vangelis Marinakis\IEEEauthorrefmark{1}}
\IEEEauthorblockA{\IEEEauthorrefmark{1} Decision Support Systems Laboratory, School of Electrical and Computer Engineering,\\ National Technical University of Athens, Athens, Greece}
\IEEEauthorblockA{\IEEEauthorrefmark{2} HOLISTIC SA, Athens, Greece}
\IEEEauthorblockA{\IEEEauthorrefmark{3} University of Thessaly, Gaiopolis Campus, 41500, Larissa, Greece}
Email: \{gkorbakis, alekidis, esarmas, jpapias, vmarinakis\}@epu.ntua.gr, \{fserepas, gstravodimos\}@holisticsa.gr, alekidis@uth.gr
}

\maketitle

\begin{abstract}
The contemporary era is marked by rapid urban growth and increasing population. A significant, and constantly growing, portion of the global population now resides in major cities, leading to escalating energy demands in urban centers. As urban population is expected to keep on expanding in the near future, the same is also expected to happen with the associated energy requirements. The situation with the continuously increasing energy demand, along with the emergence of smart grids and the capabilities that are already -or can be- offered by Home Energy Management System (HEMS), has created a lot of opportunities towards a more sustainable future, with optimized energy consumption and demand response, which leads to economic and environmental benefits, based on the actual needs of the consumers. In this paper, we begin by providing an analytical exploration of the challenges faced at both the development and deployment levels. We proceed with a thorough analysis and comparison between the abundance of devices, smart home technologies, and protocols currently used by various products. Following, aiming to blunt the currently existing challenges, we propose a reliable, flexible, and extendable architectural schema. Finally, we analyze a number of potential ways in which the data deriving from such implementations can be analyzed and leveraged, in order to produce services that offer useful insights and smart solutions towards enhanced energy efficiency. 
\end{abstract}

\begin{table}
\centering
\caption{Abbreviations}
\label{tab:abbreviations}
\begin{tabular}{|c|p{0.7\linewidth}|}
\hline
\textbf{Abbreviation} & \textbf{Meaning} \\ \hline
AI & Artificial Intelligence \\ \hline
API & Application Programming Interface \\ \hline
CHP & Combined Heat Power \\ \hline
CoAP & Constrained Application Protocol \\ \hline
DB & Database \\ \hline
DR & Demand Response \\ \hline
EU & European Union \\ \hline
EV & Electric Vehicles \\ \hline
HEMS & Home Energy Management Systems \\ \hline
HVAC & Heating, ventilation, and air conditioning \\ \hline
HTTP & Hyper Text Transfer Protocol \\ \hline
HTTPS & Hyper Text Transfer Protocol Secure \\ \hline
ICT & Information and Communication Technologies \\ \hline
IoS & iPhone operating system \\ \hline
IoT & Internet of Things \\ \hline
ML & Machine Learning \\ \hline
MQTT & Message Queuing Telemetry Transport \\ \hline
RES & Renewable Energy Sources \\ \hline
SPM & Smart Power Meter Switch \\ \hline
SSL & Secure Sockets Layer \\ \hline
TLS & Transport Layer Security \\ \hline
V2G & Vehicle to Grid \\ \hline
\end{tabular}
\end{table}


\section{Introduction}
\label{sec:intro}

In the ever-evolving landscape of continuously increasing urbanization, the concept of smart cities has emerged as a beacon of innovation and performance. Smart cities harness the power of information and communication (ICT) technologies to revolutionize city infrastructure, services, well-being, comfort in building level and resource management \cite{ur2023future,doukasmachine}. Integral to this evolution is the optimization of electricity systems, where smart technologies play a pivotal role in reducing electricity consumption \cite{saleem2019internet}, maximizing resource utilization \cite{li2022effective}, and mitigating environmental effects through forecasting or estimating energy demand \cite{sarmas2022incremental}, eventually leading to optimized energy demand response. From smart grids to smart homes, these structures epitomize the convergence of this era and sustainability in contemporary urban environments.

Addressing the rapid growth of cities and increasing populations has sparked the development of smart city initiatives across the globe \cite{bibri2018smart}. These projects harness cutting-edge technologies to enhance energy efficiency and reduce environmental impacts \cite{sarmas2022ml}. Although the focus is put on improving transportation and utility services, the management of household electricity within smart cities has not been examined exhaustively \cite{lytras2018uses}. Urban energy consumption patterns are complex and they are continuously changing, thus emphasizing the necessity for new and intelligent energy management strategies \cite{sarmas2023estimating}. As cities grow and energy demand is constantly rising, the need for optimal and user-friendly energy solutions is more evident than ever \cite{testasecca2023recent}. Old-fashioned methods fall short in addressing the problems and complexities of contemporary urban life, necessitating innovative technologies \cite{abulibdeh2024navigating}.

Moreover, the global model towards sustainability and environmental actions makes the transformation of energy use in cities a very important issue. Given that a substantial amount of greenhouse gases come from urban energy use, shifting to cleaner and more efficient energy practices is also crucial. In this context, Internet of Things (IoT) devices and Home Energy Management Systems (HEMS), supported by Artificial Intelligence (AI), stand out as key elements for improving energy efficiency and sustainability in homes \cite{merabet2021intelligent, skaloumpakas2024reshaping, mishra2023energy, sarmas2024baseline}. The expansion of IoT devices marks the beginning of a new era in home connectivity and management \cite{alam2022towards}. Homes equipped with HEMS offer improved efficiency and comfort by utilizing the latest in information and communication technologies and AI \cite{tsolkas2023dynamic}. Nonetheless, despite HEMS' vast potential, there are still obstacles to overcome in order to fully capitalize on their advantages. Challenges such as system compatibility, data security, and user friendliness need to be addressed in order to achieve widespread use and effectiveness. The ability of these systems to meet the diverse requirements also plays a crucial role in their deployment.

Amidst these challenges, this paper aims to provide an overview focusing on the intersection of smart cities and residential energy management systems, with an emphasis on IoT-enabled HEMS. By examining current trends, challenges, and possibilities, we aim to provide insights into the transformative potential of integrating smart home technologies into the fabric of city power ecosystems. Through a comprehensive review of current infrastructure and technical solutions, we endeavor to elucidate the contribution of modern HEMS in shaping the future of carbon-neutral urban dwellings. Concretely, the contributions of this paper are as follows:

\begin{compactitem}
    \item Analysis of the main existing challenges in HEMS systems, including device/communication protocol interoperability, data management, security and privacy as well as the various stakeholder involvement. 
    \item Categorization and selection criteria for the available HEMS devices in the energy market.  
    \item Proposed structured and layered design architecture for integrating HEMS devices and services towards the development of a smart city ecosystem.
\end{compactitem}

The rest of the paper is organized as follows. Section \ref{sec:challenges} provides an overview of the main challenges in HEMS systems. Section \ref{sec:devices_cat} introduces criteria for device characterization and an Section \ref{sec:devices_eval} presents evaluation of some widely used devices on the market. Following, Section \ref{sec:solution} provides an overview of the proposed solution for integrating all the devices and technologies of HEMS systems. A discussion of services that can be developed as future work, based on the proposed solution and the data that can be collected and leveraged, follows in Section \ref{sec:discussion}. Finally, Section \ref{sec:conclusion} provides the concluding remarks.

\section{Challenges}
\label{sec:challenges}

The addition of HEMS systems within the smart grid is a significant step towards the transition to smart cities \cite{LOTFI2022118241}, yet, both the development of such systems and their smooth integration come with a plethora of challenges that vary across several domains.

Firstly, achieving seamless interoperability among a wide range of diverse and heterogeneous devices \cite{BEAUDIN2015318}, smart meters \cite{rossello2011towards}, and systems is a complex task. It involves the integration of different development technologies, semantics \cite{10.1007/978-3-030-97652-1_40} \cite{de2022semantic}, input/output formats, heterogeneous data, and communication protocols, such as Message Queuing Telemetry Transport (MQTT), Constrained Application Protocol (CoAP), Zigbee, etc. \cite{lee2016interoperability}. On top of these, proper resource management \cite{shafik2020internet} emerges as a critical concern. Overcoming these hurdles often represents the primary and most significant obstacle in such endeavors, underlining the need for innovative and, primarily, flexible solutions.

Furthermore, efficiently managing the vast data streams produced by HEMS also poses a significant challenge \cite{JIA2019643}. Ensuring fine-grained access control, privacy, and cyber-security \cite{6345767} \cite{7028872} is also of crucial importance, given that this data often contain sensitive information about individuals, such as energy consumption metrics and usage patterns. Thus, this data must not be vulnerable to unauthorized access. Instead, this data should only be accessed by authorized users and system developers for analysis and insights extraction. HEMS face a range of security concerns \cite{jabraeil2020iot}, including malware threats, the misuse of default or simple device credentials, particular vulnerabilities within software and hardware, and flaws in network security, each contributing to potential breaches of privacy, system integrity, and overall smart home functionality \cite{article}.

The challenges described above, primarily technical constraints in developing HEMS, are essential to ensure they achieve the necessary resilience and reliability \cite{hussain2021demand}. However, beyond these technical aspects, the integration of HEMS with smart grids introduces further complexities, including equipment-specific challenges and the need for consistent updates and maintenance \cite{depuru2011smart}. Such requirements emphasize the importance of device durability and adaptability in maintaining system effectiveness. 

One crucial factor towards the successful integration of HEMS into the smart cities network, is the collaboration needed between the various stakeholders, which include city governments, utility providers, technology vendors and, of course, community members. Effective stakeholder engagement emerges as a point of significant importance for receiving the needed support, gaining the needed motivation, fostering collaboration and addressing the concerns that can arise in the process \cite{doi:10.1177/1420326X211001698}. Additionally, the escalating energy demands in rapidly expanding urban areas also pose a significant challenge, since they keep widening the potential gap between energy supply and demand \cite{PANDIYAN2023648}, creating a pressing issue that requires innovative solutions.

The technological implementation of HEMS, as well as their subsequent integration into the smart cities grid is a big challenge. Without the proper user engagement, the benefits of such an integration cannot be unlocked, since user engagement and participation is essential for the success of this endeavor \cite{tadili2019citizen}. Therefore, providing proper motivation in order to encourage active involvement is highly necessary in order to maximize the potential outcomes.

Integrating HEMS within smart grids plays a crucial role towards advancing to smart cities. However, as described above, there are a variety of challenges and considerations across numerous domains that require meticulous attention and efficient, well-crafted solutions to be overcome.

\section{HEMS Device categorization}
\label{sec:devices_cat}
HEMS are complex implementations and the devices deployed on the field to realize them can be separated in three main categories, namely Smart Meters, Sensing Devices and Control Devices. 

Smart Meters are devices used to measure energy consumption, on device, apartment, or building level, including real-time measurement capability, and transmit the collected data \cite{mahapatra2022home} for future analysis and research. Sensing Devices are sensors deployed within residencies or buildings aiming to measure, gather and transmit data regarding various environmental conditions, such as temperature, humidity, occupancy and light intensity \cite{kim2016hems}, eventually transmitting, as well, the collected data for analysis and patterns recognition. Lastly, Control Devices empower users to manage and control the operation of their connected appliances. Some examples are smart thermostats and smart plugs, while it is worth mentioning that several control devices also support metering functionalities.

As mentioned in the previous section, the abundance of devices and smart home technologies is an undoubted challenge towards implementing a Home Energy Management System, since achieving communication between different smart devices, which communicate over different protocols and follow varying semantics is an inherently complex task. Making the decision on which devices to leverage in order to achieve the desired result is a crucial step that must be taken at the very start of each undertaking, as it is a choice that will affect the whole architecture of the future implementation. This decision is a vital factor towards seamless integration, reliability, maintainability, and extendability.

To begin with, taking into consideration the connectivity mechanisms and protocols (e.g., MQTT, CoAP, or HTTP requests) employed by each device is crucial, as it directly impacts how data is exchanged across the system's nodes. The breadth of measurements each device can undertake is also a significant criterion, as it obviously determines the final capabilities and functionalities of the HEMS that will be implemented. From measuring and monitoring temperature and humidity, up to tracking energy consumption, there is a wide -and continuously expanding- range of possible measurements, that can offer invaluable, exploitable insights and detailed data after proper analysis. 

At the same time, the parameterization capabilities and the devices' versatility play a key role in device selection, as these characteristics affect the resulting capability of providing tailored configurations and flexible solutions. 

Security and data privacy stand as paramount considerations within the energy sector, given the sensitivity of energy consumption \cite{kormpakis2023energy} and residence devices' data. Unauthorized access, as a result of privacy breaches, could potentially compromise the safety of the occupants. Thus, ensuring robust and secure devices and data exchange mechanisms remains an undeniable priority in the implementation of HEMS. 

Moreover, the concept of portability is an important aspect as it affects whether a solution can be easily adopted in a household, while openness, meaning whether the devices' functionalities and interfaces are open and accessible or proprietary systems, must be also considered during the design of a HEMS, as it can affect how flexible and customizable the final implementation will be. 

Finally, the cost of the devices remains a practical consideration and it is a factor affecting a lot of aspects of the final implementation, such as scalability and longevity. 

Summarizing, the selection of devices for a HEMS requires careful consideration of various criteria, including connectivity mechanisms, measurement capabilities, parameterization possibilities, security, portability, openness, and devices' cost. Each one of the above-mentioned criteria plays a crucial role in shaping the architecture, functionality list, and overall viability of the HEMS implementation.

\section{HEMS Device Evaluation}
\label{sec:devices_eval}

By considering the aforementioned criteria, 
we proceeded reviewing several smart devices of the market, aiming to identify the most suitable ones for designing a flexible, reliable and extendable HEMS architecture that can be integrated in a smart city ecosystem. Taking into consideration the length restrictions of a conference paper, the review provided below is not exhaustive, however it displays a variety of well-known and widely used available solutions, which cover the whole spectrum of the desired functionalities of the final implementation. 

Initially, Shelly devices \footnote{\url{https://www.shelly.com/}}, including Shelly EM, 3EM and Smart Plug S, offer a diversity of functionalities for energy management and monitoring. Shelly EM and 3EM devices provide energy monitoring and real-time measurements of energy consumption at both device and residence levels, while they also incorporate features for controlling connected devices to optimize energy usage. Shelly Smart Plug S, on the other hand, offers remote control capabilities for any plugged-in devices, enabling users to remotely control the functioning of their devices using user interfaces which are accessible via the web. Supporting a diverse range of connectivity mechanisms like Wi-Fi, MQTT, and CoAP, these devices ensure seamless communication within smart home networks. These devices provide real-time energy consumption measurements on device and residence level, as well as device controlling capabilities. They also incorporate a high-level of parameterization capabilities and versatility, offering access to various settings and personalized preferences, as well as an open, yet secure, Application Programming Interface (API) and support for third-party integration, allowing customized configurations and adaptable solutions. Notably, their affordability enhances the Shelly devices' accessibility to a wide spectrum of consumers.

The SONOFF devices \footnote{\url{https://sonoff.tech/}}, such as Smart Power Meter Switch (SPM-MAIN), BASICR4, and RFR2, represent solutions for home energy management within the SONOFF ecosystem. Remote control on devices and real-time energy consumption measurements are offered to the end-users via these devices, which can be seamlessly integrated into existing home networks via Wi-Fi, providing the highly desired convenience and flexibility in household appliances management. Other communication protocols, such as MQTT are also supported, enhancing interoperability with different smart home devices or systems. Data privacy and security are principles followed by design in SONOFF devices, with encryption protocols such as SSL/TLS being used to encrypt data transmission between the devices, while a data privacy policy, ensuring that usage and storage of data happens in aggregated and anonymized form so that it is not associated with individual end users and does not include personal information, is declared in the official documentation \footnote{\url{https://sonoff.tech/privacy-policy/}}. Notably, a number of SONOFF devices also expose an Open API, enabling advanced users to customize settings and preferences according to their specific needs, while they are also quite affordable. 

Additionally, Xiaomi offers a wide range of smart home devices, with various functionalities that can be integrated into a HEMS. These functionalities include a variety of measurements, offered by devices like the Xiaomi Mi Temperature and Humidity Sensor which fosters indoor conditions quality and sensors like window and door sensors. Additionally, various other practical functionalities that facilitate, such as facilitating users' interaction and enabling remote control of their devices, are provided by devices like the Xiaomi Mi Smart Plug and Smart LED Bulbs. Using primarily Wi-Fi, they can support seamless integration within their ecosystem, which also uses proprietary protocols for data exchange. Parameterization capabilities differ among the different devices, offering varying levels of customization, while privacy policy procedures are followed \footnote{\url{https://g.home.mi.com/views/user-terms.html?locale=en&type=userPrivacy}} and robust security mechanisms are implemented, as declared in the official documentation \footnote{\url{https://trust.mi.com/pdf/MIUI_Security_White_Paper_EN_May_2021.pdf}}, to ensure data privacy. While Xiaomi offers capabilities for developers to integrate their devices with third-party platforms to some extent, the level of openness and compatibility is not as extensive as with other ecosystems, as the proprietary application is primarily used as an integration point of all the devices connected to the ecosystem. Regarding their cost, Xiaomi devices are generally considered to be affordable, which enhances their accessibility to a wide range of users.

Netatmo is another company that offers a diverse range of smart home solutions, primarily designed to enhance energy management and monitoring. This is possible via a number of devices, like the Netatmo Smart Thermostat and Weather Station. Indicatively, the Smart Thermostat enables users to remotely control their heating system and adjust temperature, while the Weather Station offers real-time data on indoor and outdoor conditions, providing measurements for temperature, humidity, air quality and energy consumption. Leveraging Wi-Fi connectivity and integration with widely used smart home platforms, Netatmo devices also offer seamless communication and interoperability within a smart home ecosystem. In terms of security, Netatmo prioritizes security with robust encryption protocols and regular firmware updates to safeguard sensitive data, while being aligned with European data privacy regulations \footnote{\url{https://www.netatmo.com/en-gb/security-guide/protection-of-personal-data}}. Additionally, with the availability of an open Application Programming Interface, Netatmo devices allow data access and data retrieval, facilitating the integration with third-party platforms and the creation of custom applications. Finally, Netatmo devices are considered as affordable. 

Further energy metering devices are the smart meters that are part of the Advanced Metering Infrastructure \cite{lekidis2024towards} and offering electricity consumption monitoring as well as real-time measurement availability for both utility operators and household customers. Smart meters are manufactured by various manufacturers including Landis+Gyr\footnote{\url{https://www.landisgyr.com/}}, Sagemcom\footnote{\url{https://www.sagemcom.com/en}}, Honeywell's Elster\footnote{\url{https://process.honeywell.com/us/en/site/elster-instromet}} and Hager \footnote{\url{https://hager.com/intl-en}} amongst the well-known ones. The protocols that are used for communication are DLMS/COSEM\footnote{\url{https://www.dlms.com/}}, Modbus\footnote{\url{https://modbus.org/}} and M-Bus \footnote{\url{https://m-bus.com/}} and they ensure high compatibility with various smart home devices, networks and systems. 

One example of energy metering devices, is the Hager HTG411H. The focus of this device is on providing accurate energy consumption measurements, contributing to effective energy management within buildings, by providing access to local dashboards with various visualizations and graphics displayed with embedded web-pages, commissioning reports and file export capability. They also incorporate robust security mechanisms, user management implementation and encryption techniques (e.g., communication over HTTPS) to ensure data privacy. The cost of these devices is significantly higher than the previously mentioned ones, while they typically operate within a closed ecosystem, meaning that their functionalities and interfaces are proprietary and specifically designed to work within the company's system architecture.

Given all these challenges and different criteria for selecting devices, we present in the following section an architecture for the optimal design of HEMS systems as well as enabling their integration with further systems towards the development of a smart city ecosystem. 

\section{Proposed HEMS Solution}
\label{sec:solution}

\begin{figure*}[!htbp]
\centering
\includegraphics[width=0.8\textwidth]{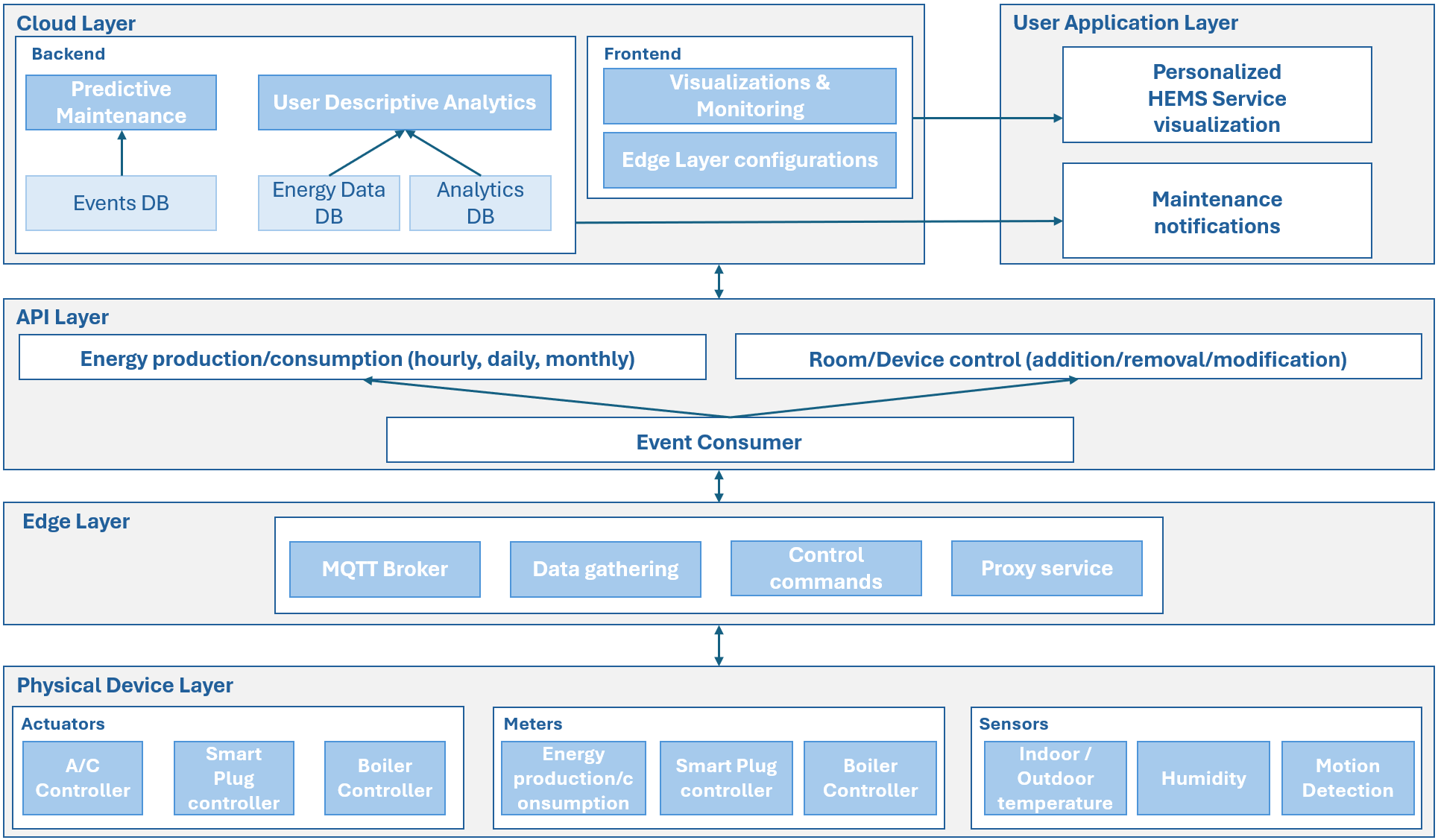}
\caption{Proposed HEMS system architecture}
\label{fig:architecture}
\end{figure*}

The starting point for designing a robust HEMS solution lies in addressing efficiently the challenges of Section \ref{sec:challenges} as well as integrating the device categories that were presented in Section \ref{sec:devices_cat}. Considering that there are different categories of devices, the architecture shall be structured and layered in order to also reflect the data flow, as depicted for example from the Purdue Model architecture\footnote{\url{https://www.energy.gov/sites/default/files/2022-10/Infra\_Topic\_Paper\_4-14\_FINAL.pdf}}. 

As an outcome, we derived the layered architecture of Fig. \ref{fig:architecture} for HEMS systems, which is composed from the following layers (presented from bottom to top based on data flow):

\begin{compactenum}
    \item \textbf{Physical Device Layer}: This layer includes the devices within a household, which require energy use to operate. Specifically, the devices include sensors, collecting environmental data which is crucial for intelligent control including a) indoor/outdoor temperature monitoring for energy-efficient heating or cooling, b) humidity for measuring moisture levels which can affect heating and cooling needs and c) motion detection for optimizing lighting and HVAC usage. Additionally, there are also meters that provide measurements of the energy production/consumption from household devices (e.g., HVAC, EV chargers). Finally, there are also device controllers for managing the individual devices, such as the A/C Controller for managing the air conditioning systems, the Smart Plug Controller that allows controlling (turning ON or OFF) over smart plugs (e.g., Shelly plugs), as well as the boiler controller that manages the operation of the boiler for heating purposes. Shelly devices are a preferable option, since they offer increased customization and extendability potential, as described in Section \ref{sec:devices_eval}. On the same time, they provide real-time measurements and remote control capabilities via a dedicated backend application that is further explained below. 
    \item \textbf{Edge Layer}: This layer implements edge processing and intelligence at the HEMS device level. The core component in this layer is the \textit{Home Gateway}, which acts as a central hub for analyzing the received data from different sensors as well as for performing computations that may lead into control actions on the actuators. A device that can be used to offer these functionalities, is the Raspberry Pi \footnote{\url{https://www.raspberrypi.org/}}. The device cost is considerably lower than further gateway devices on the market, while it also offers high computation power to support edge-based data processing algorithms. Additionally, it offers a proxy service for 1) parsing/translating data and requests from one communication protocol to another and 2) forwarding the data from the sensors/actuators towards the Cloud layer. The opposite flow is also possible, i.e., parsing/translating and then forwarding requests from Cloud layer services towards the computing servers. Furthermore, the Edge Layer includes the Event Consumer component, which listens for events or data from both the cloud layer and physical devices to take appropriate action. 
    \item \textbf{API Layer}: This layer acts as the middleware facilitating data exchange from the edge towards the Cloud layer. Specifically, it allows the exchange of energy production/consumption data for different time frames (hourly, daily, weekly, monthly, yearly). Additionally, it provides an interface to control and configure smart devices and room settings, including adding, removing, or modifying device settings.
    \item \textbf{Cloud Layer}: This layer serves a central processing/orchestration layer and includes the components:
        \begin{compactitem}
        \item \textit{Back-End}: It is composed initially of the predictive maintenance service which uses historical and real-time data to predict potential equipment failures or maintenance service. Moreover, the Events DB is a database for logging events, which can include sensor data, user actions, and system alerts.
        \item \textit{Energy Data DB}: Stores data related to energy consumption and production. A prominent database in this category is the Timescale DB \footnote{\url{https://www.timescale.com/}} which is optimal for storing the energy-based timeseries data.
        \item \textit{Analytics DB}: Contains data used for analytical purposes, like user behavior patterns or energy usage trends.
        \item \textit{Front-End}: This component contains graphical interfaces (i.e., Visualizations in Fig. \ref{fig:architecture}) that display energy consumption patterns on device and household level, disaggregation metrics, system status, and predictive maintenance information in an understandable format for the user. Moreover, it also provides settings and controls that allow users to manage and configure the edge devices and sensors (i.e., Edge Layer Configurations in Fig. \ref{fig:architecture}). For the implementation of the Front-End application, a wide variety of technologies can be utilized. For example, React \footnote{\url{https://react.dev/}} is a state-of-the-art, free, open-source JavaScript \footnote{\url{https://www.javascript.com/}} library used for the creation of dynamic user interface components on websites and apps, without the need for page reloads. In order to present visualization and advanced monitoring capabilities, a variety of libraries can be also used. A few examples of JavaScript-based libraries, that can be integrated in such a dashboard, are Chart.js \footnote{\url{https://www.chartjs.org/}} and amCharts \footnote{\url{https://www.amcharts.com/}}.
        \end{compactitem}
    \item \textbf{User Application Layer}: This layer is depicted in the same level as the Cloud as the user has a view of the services in a mobile (i.e., Android, IoS) application, which can be leveraged to perform control actions towards the HEMS system or the individual devices.
\end{compactenum}

\section{Potential HEMS services}
\label{sec:discussion}

As HEMS systems are gradually integrated within smart cities, there is a vast potential for the development of smart services aimed at enhancing building performance and optimizing energy savings. Such services could offer valuable insights into energy consumption behaviors, empowering stakeholders to make informed decisions regarding energy usage and management. The services that are described below are based on the proposed HEMS architecture of Section \ref{sec:solution} and link to the different introduced layers of Fig. \ref{fig:architecture}. 

Initially, detailed insights into energy usage patterns within the households may be offered as a potential service. This involves the utilization of historical and near-real-time sensor data for environmental conditions and energy consumption from temperature, humidity, and occupancy sensors as well as smart meters (Physical Device Layer). The gathered data are processed in the Edge Layer, where a Home Gateway, acting as the central hub, performs initial data filtering, aggregation, and analytics to extract relevant insights locally. Additionally, edge processing capabilities would enable real-time monitoring of energy usage patterns within the building. Once processed at the edge, the data can be also provided to the Cloud Layer through the API Layer where advanced analytics and visualizations may be offered to monitor and pinpoint energy-intensive areas and system flows within the household. 

Moreover, another potential service could support household occupants in identifying faults and inefficiencies in HEMS systems, such as equipment malfunctions, through the analysis of performance indicators from the devices of the Physical Device Layer. In such service, ML-based fault detection and diagnostics algorithms within the Edge Layer shall be leveraged to identify anomalous patterns. Once anomalies are detected, the data would be transferred to the Cloud Layer through the API Layer to 1) investigate on potential equipment malfunctions/suboptimal energy performance and 2) perform mitigation actions as a part of decision-making. Mitigation actions may involve the use of control devices such as smart thermostats and smart plugs for performing preventive maintenance actions, such as adjusting HVAC settings or turning off malfunctioning appliances. This service may aid towards reducing downtime and improving the overall system reliability. 

A third service concerns the availability of personalized recommendations to household occupants. In this service, the Cloud Layer would employ ML models to produce tailored recommendations for optimizing energy efficiency that would be communicated to users through the User Application Layer. The User Application Layer serves as the interface for household occupants to receive and act upon the recommendations, providing actionable insights for optimizing energy usage. Through dedicated mobile applications, occupants shall be able to view and perform the recommended actions in real-time, enhancing their proactive engagement in energy-saving behaviors. Such actions may involve adjusting thermostat settings, optimizing lighting usage, and scheduling appliance usage more efficiently. Lighting systems in this case are controlled by smart switches, allowing for dimming or switching off lights in unoccupied rooms, while smart thermostats could regulate HVAC systems, adjusting temperatures based on the presence or absence of occupants. 

Furthermore, a service centered around electrification of energy and thermal demand could be built based on the EU 2050 long-term strategy \footnote{\url{https://climate.ec.europa.eu/eu-action/climate-strategies-targets/2050-long-term-strategy\_en}} technologies, integrating Renewable Energy Sources (RES) as well as thermal/cooling systems as heat-pumps. Heat pumps constitute a major pillar of the EU strategy, as they provide for heating and cooling needs using electricity rather than direct fossil fuel combustion. Moreover, this service would allow conducting Demand-Response (DR) scenarios \cite{MICHALAKOPOULOS2024122943}, which may also involve the integration of Electric Vehicles (EVs) and associated EV chargers as an extensive electricity load consumer \cite{guzman2023intelligent} in the Physical Device Layer of the HEMS architecture. Such integration will connect HEMS with smart mobility services, in order to form an interconnected smart city ecosystem. Additionally, DR scenarios will allow balancing demand and supply in the electricity grid as well as reduction of energy costs. For this service, DR algorithms incorporating the loads from RES, heat pumps and EVs have to be executed close to the user to solve electricity grid issues (e.g., blackouts, brownouts) locally near the location they occur. For instance, the DR algorithms may 1) respond to demand peaks by curtailing energy usage or 2) activate energy storage systems during excess production from RES, in order to consume it when solar or wind sources of power for RES are unavailable. For this service, the Home Gateway in the Edge Layer is an optimal choice for the deployment and execution of DR algorithms. 

Another category of HEMS services are related to flexibility, which links to the ability of a electricity grid to maintain reliable operations in the face of rapid and unpredictable changes in supply or demand. Specifically, HEMS may leverage RES and energy storage systems, to allow the optimal matching between the demand and the supply of electricity. Furthermore, since EVs can serve as mobile energy storage units, flexibility scenarios may incorporate Vehicle-to-Grid (V2G) technologies \cite{bibak2021comprehensive} for allowing the discharge of EV batteries to feed energy back into the grid during peak times. Moreover, HEMS may shift high-energy-consuming appliances to off-peak hours when electricity demand is lower as well as potentially cheaper. This may include moving the operation of heat pumps, electric water heaters, and EV charging to timeframes with excess RES energy production. Subsequently, increased availability of the electrical grid would be possible and grid issues (e.g., blackouts, brownouts) can be avoided similarly to the DR service. For the flexibility service, the Edge Layer shall employ dedicated algorithms to 1) perform load shifting and 2) management of energy storage systems as well as 3) coordination with grid operators for taking into account grid signals and tariff information to provide incentives for encouraging energy-saving and contributing to grid flexibility. Furthermore, by monitoring energy market dynamics and regulatory developments, the Cloud Layer would enable strategic decision-making regarding energy procurement and grid infrastructure investments. Through predictive modeling, the Cloud Layer would anticipate future grid conditions and recommend proactive measures to enable additional flexibility opportunities. 

As HEMS systems integrate into smart cities, key stakeholders have the opportunity to benefit in various ways. Homeowners and residents enjoy improved energy efficiency, reduced electricity consumption costs, and enhanced comfort (including devices' malfunction notifications) through personalized services. Utility companies gain from better grid stability and reduced peak demand. Government and regulatory bodies achieve energy efficiency and carbon reduction goals, setting standards and providing incentives. City planners and smart city developers use HEMS data to enhance urban planning and infrastructure integration.

\section{Conclusion}
\label{sec:conclusion}
This paper provides an overview of HEMS systems and their potential integration within the smart city ecosystem, aiming to increase citizen comfort and environmental sustainability amidst an era characterized by constantly growing population and urbanization. This is achieved by initially identifying the challenges of achieving interoperability between diverse devices, managing vast volumes \cite{kormpakis2022advanced} of heterogeneous data, and addressing security concerns. 
Additionally, criteria for selecting the appropriate devices tailored to each HEMS are presented, along with technologies, and protocols that are currently used in existing HEMS systems. Based on the selection criteria and by considering the system challenges we propose a new structured and layered HEMS architecture, allowing 1) to provide various services at different domains as well as 2) integrate with further system categories, such as the smart mobility systems, towards forming a smart city ecosystem. Following on the proposed HEMS architecture, we provide a list of services that may be used by household occupants, facility managers as well as grid operators, including user descriptive analytics, predictive maintenance, DR as well as flexibility algorithms. 

In summary, this paper delves into the hurdles of integrating HEMS within the smart city ecosystem, but also underscores the importance of stakeholder engagement and collaboration to initially overcome these challenges. Subsequently, such collaboration will also aim to maximize the integration benefits and ensure the effective deployment of the planned integration. As a next steps of our work, we currently focus on the development of the presented HEMS architectural framework as well as in the implementation of the associated services within the BuildON and META BUILD Horizon Europe projects. Specifically, we aim to explore how the framework's capabilities can be harnessed and expanded to yield even more insightful outcomes, including AI outputs for optimized energy consumption, demand-supply alignment, and indoor comfort enhancement for occupants within the residential sector. Additionally, we plan to ensure the proposed framework's scalability by deploying it in large-scale environments that is required to allow extensive testing for ensuring the adequate scalability of an actual smart city integration.

\section*{Acknowledgment}
This work has been co-funded by the European Union under the EU Programme Horizon-CL5-2022-D4-01-03 within the BuildON project, titled 'Affordable and digital solutions to Build the next generation of smart EU buildings', with Grant Agreement Number: 101104141. Additionally, it received funding under the META BUILD project, 'Powering the METAmorphosis of BUILDings towards a decarbonised and sustainable energy system', supported by grant agreement No. 101138373, within the Horizon Europe Research and Innovation Programme.

\printbibliography 

\end{document}